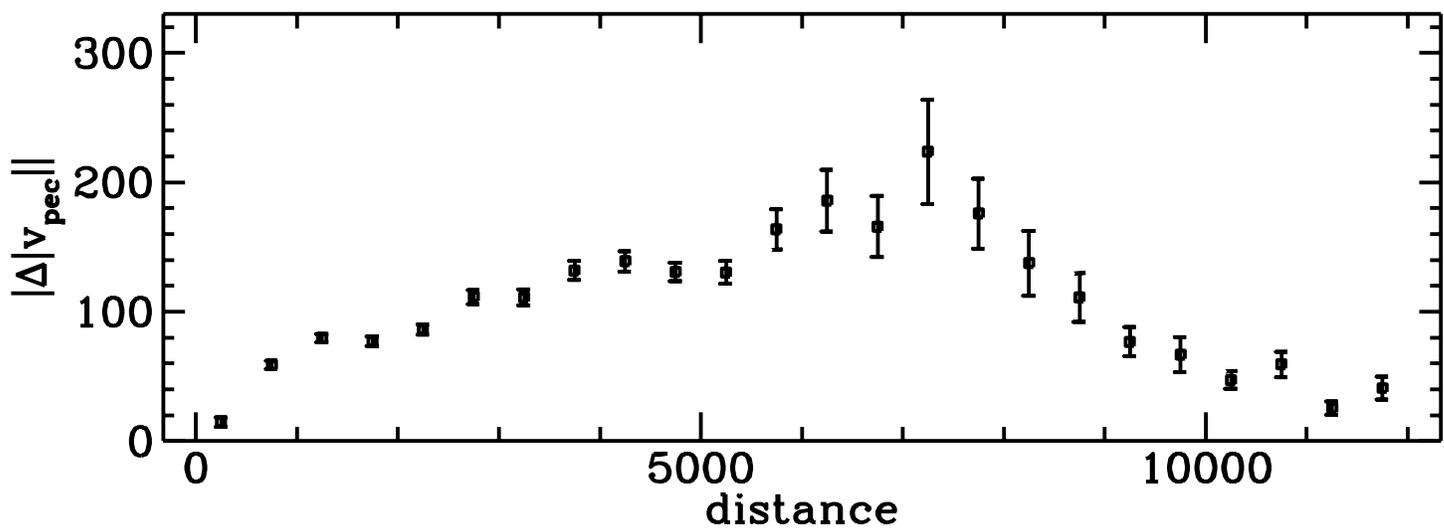
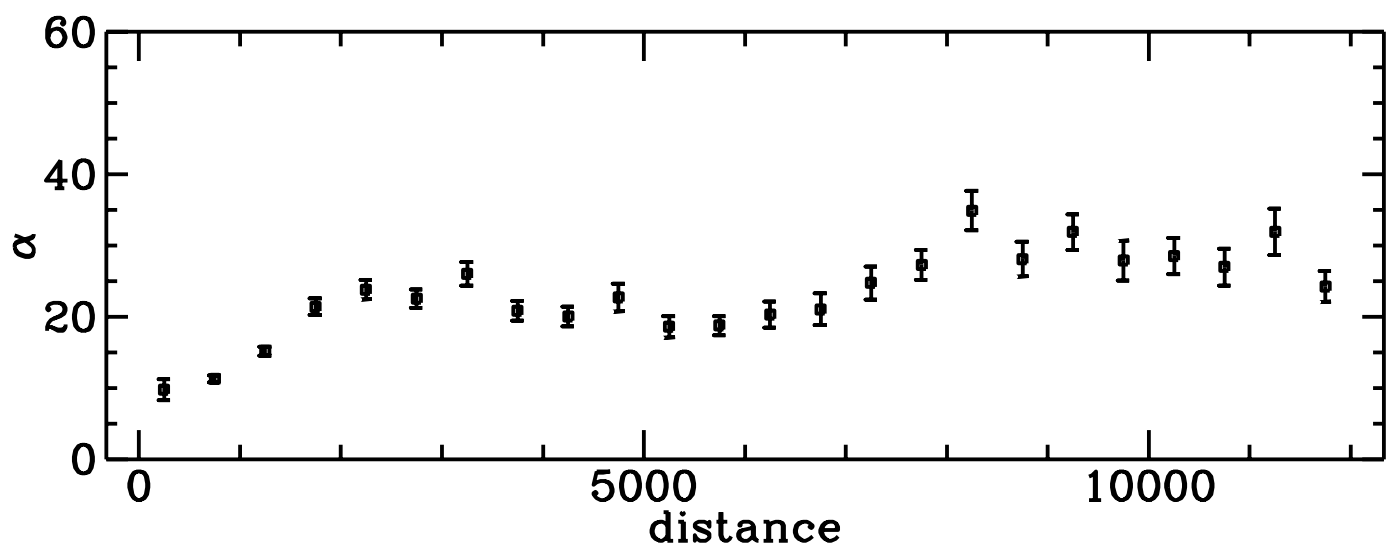
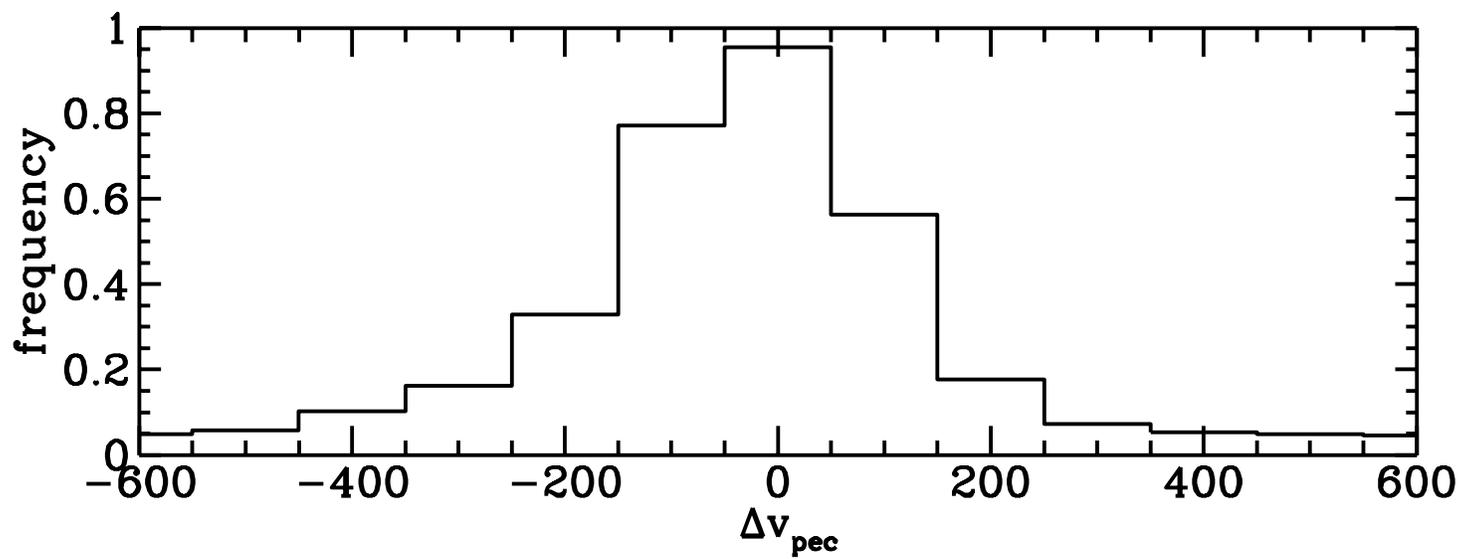

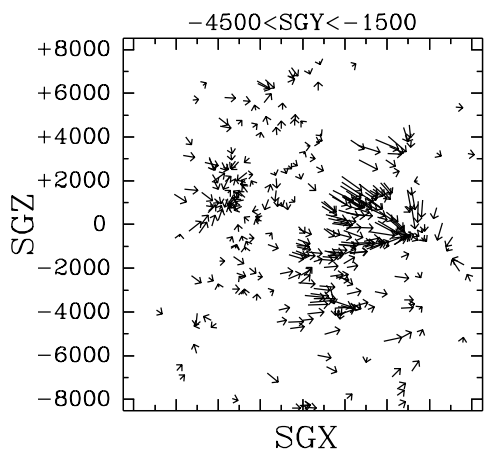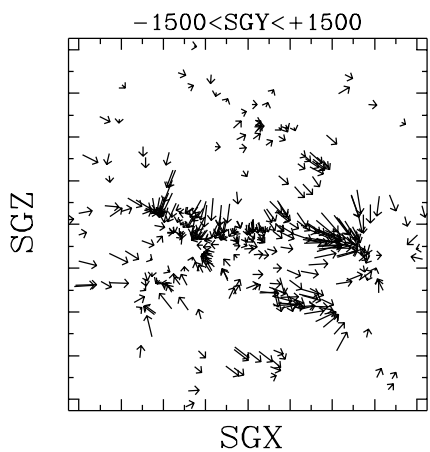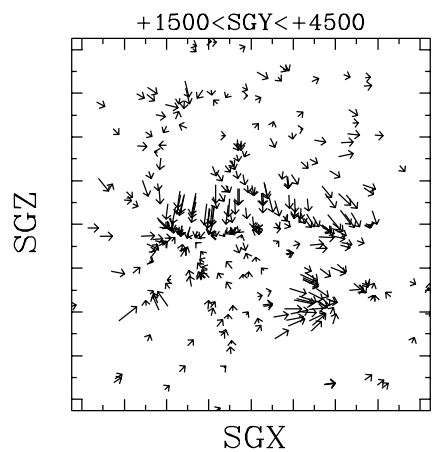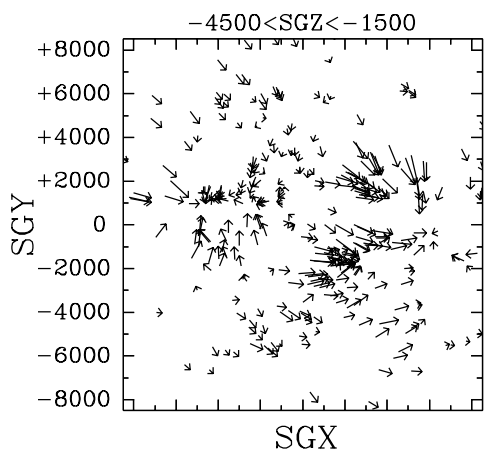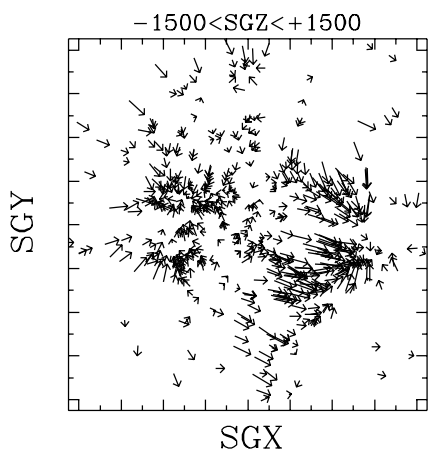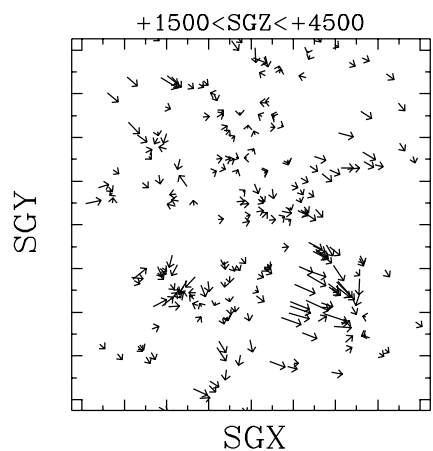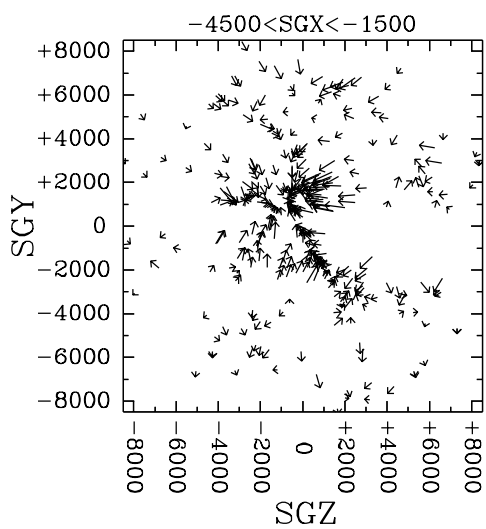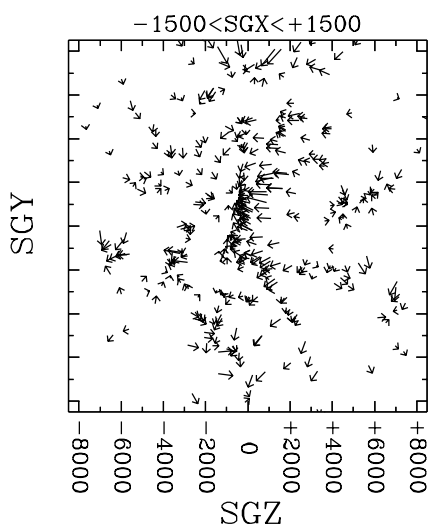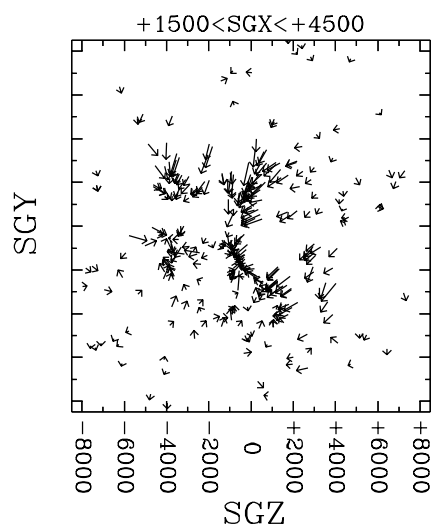

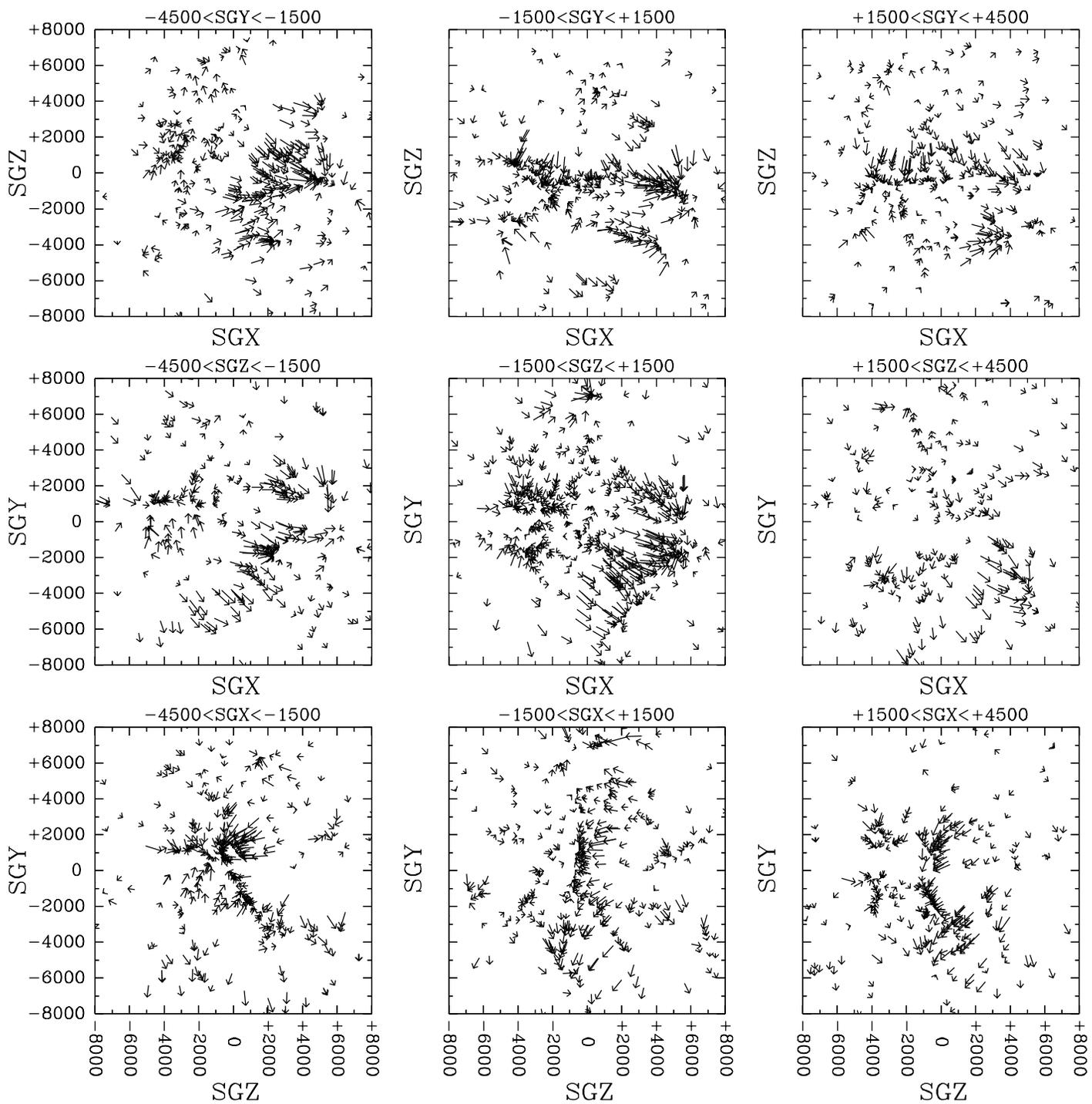

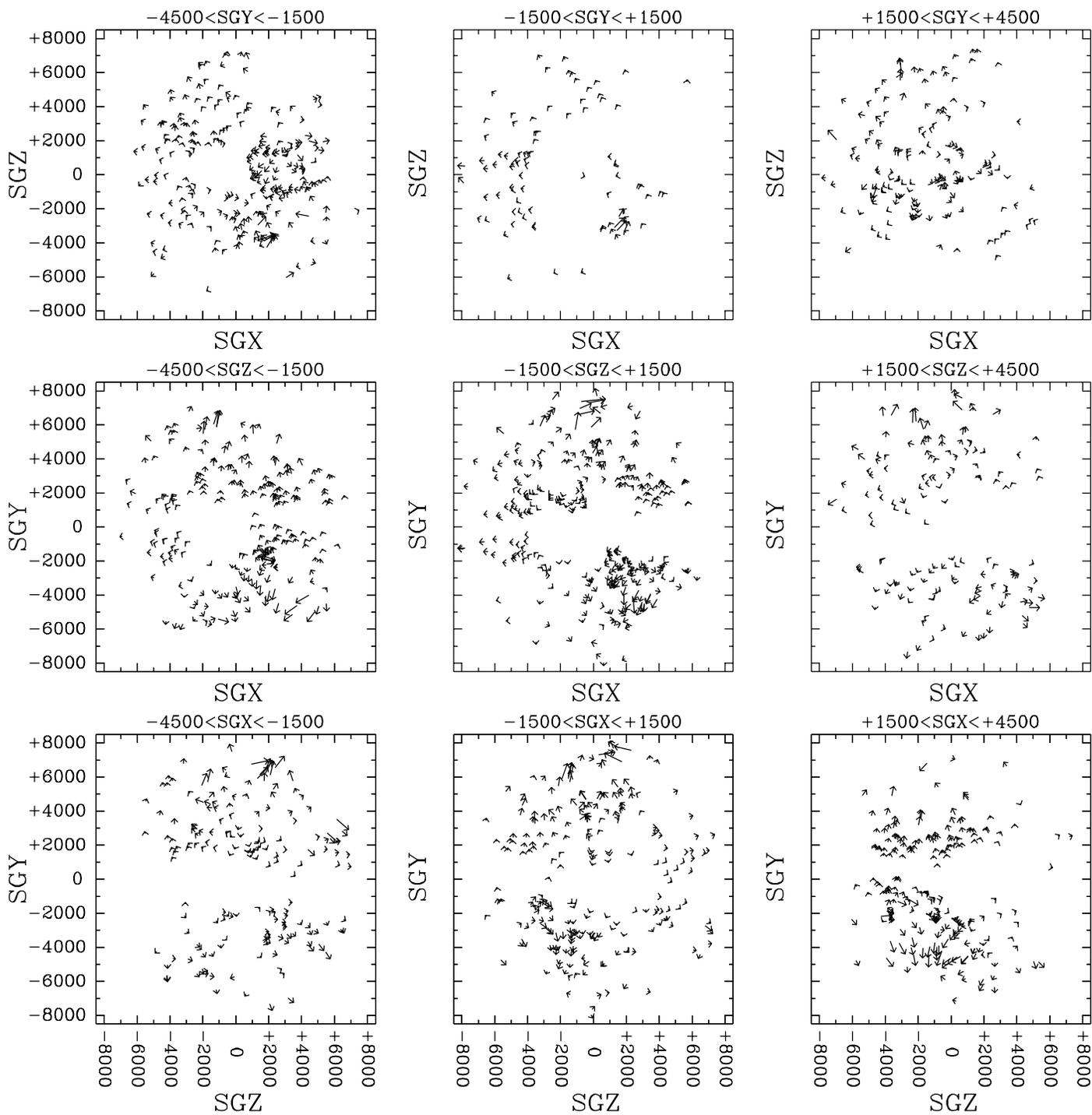

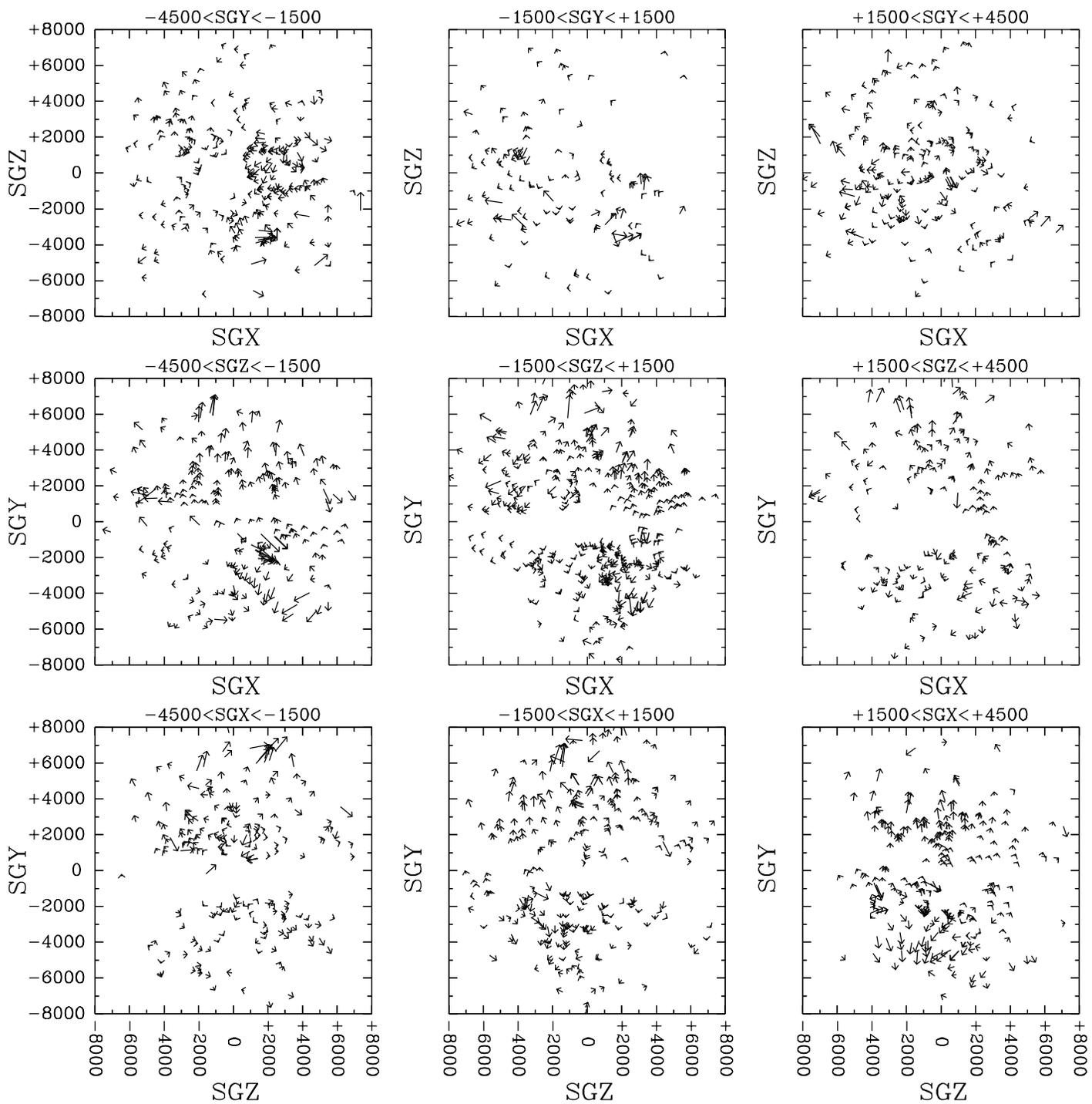

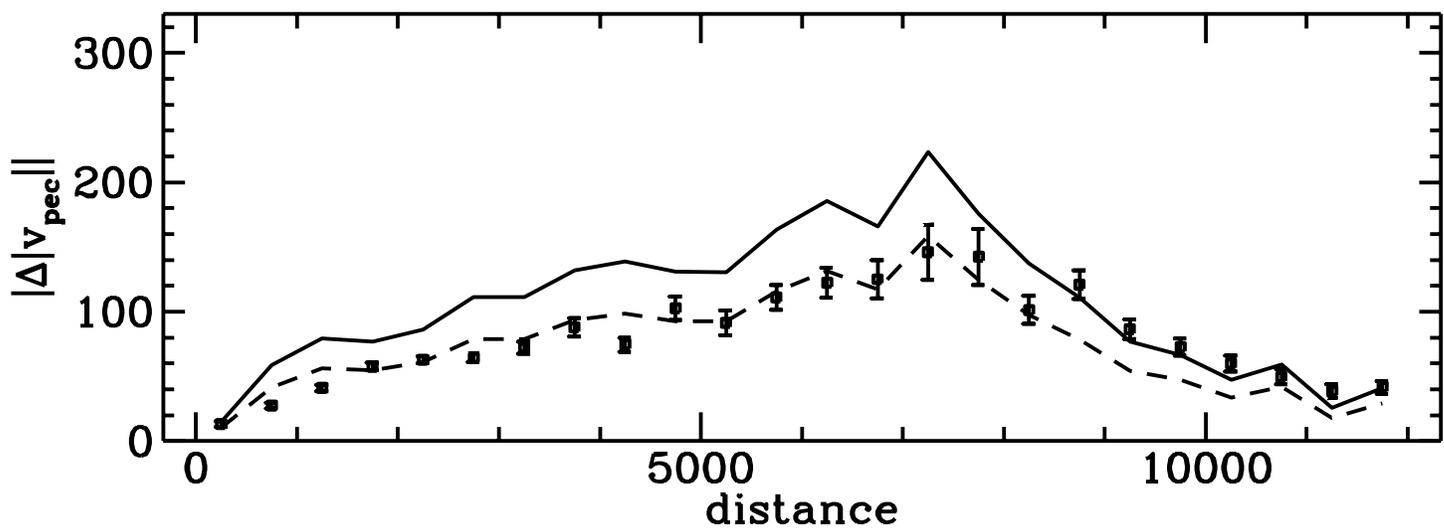
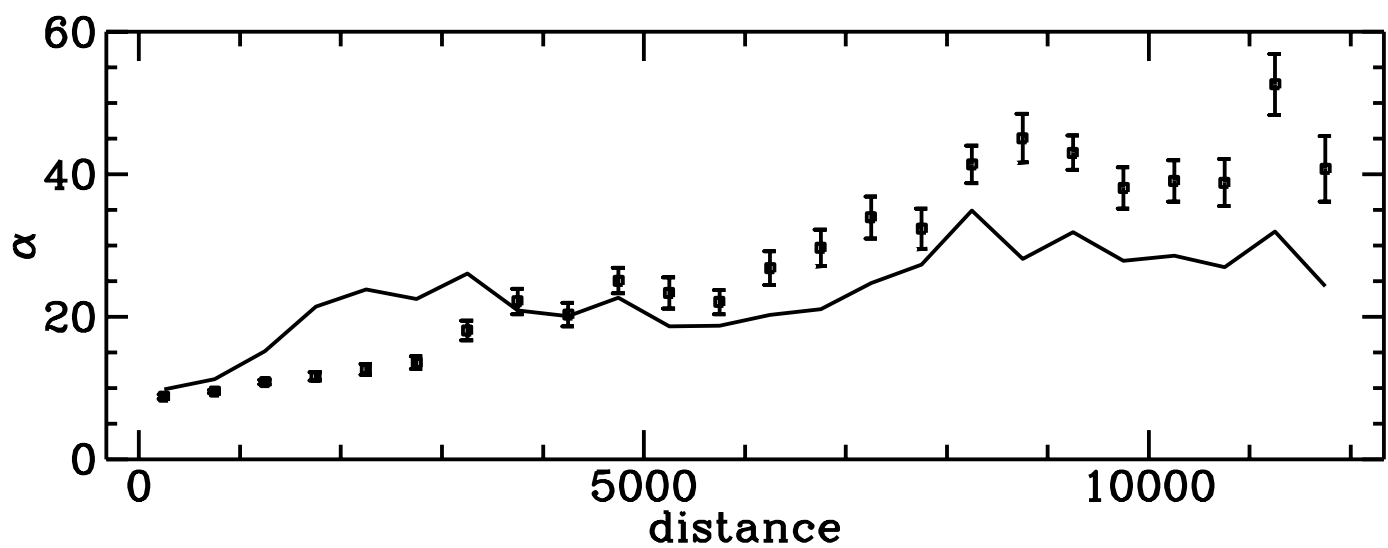
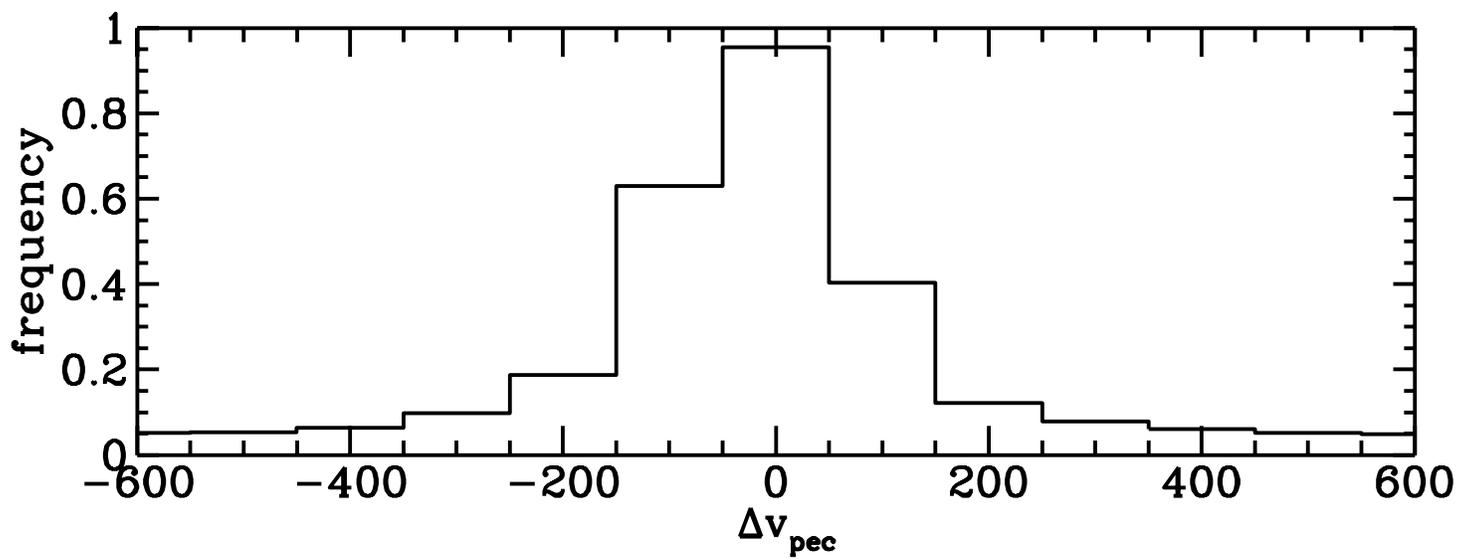

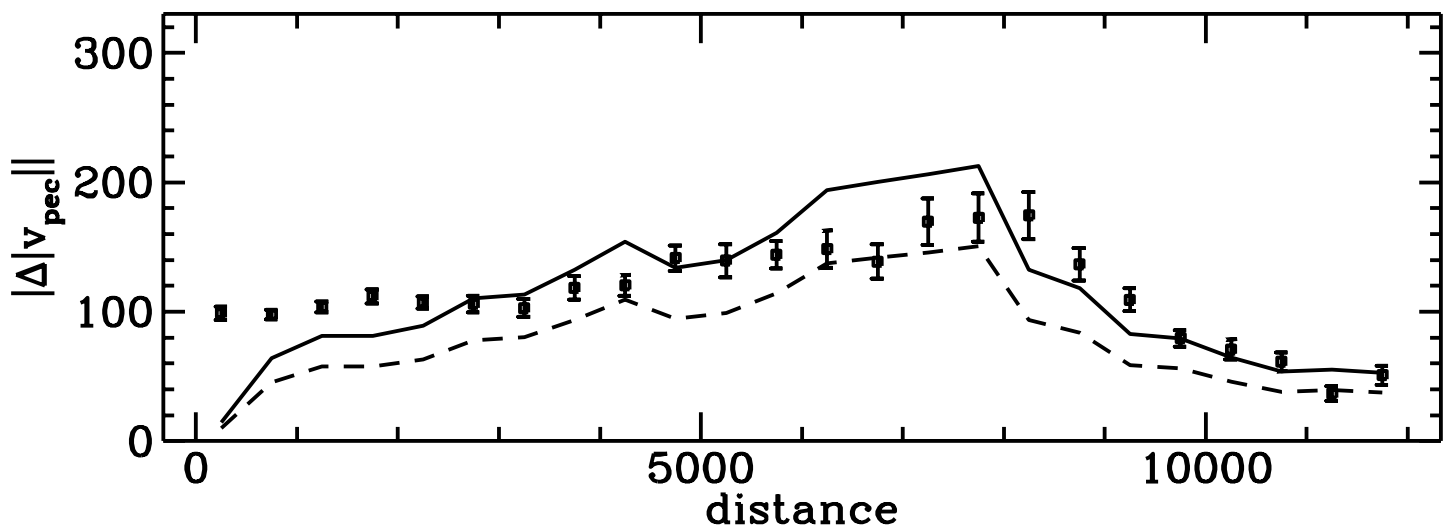
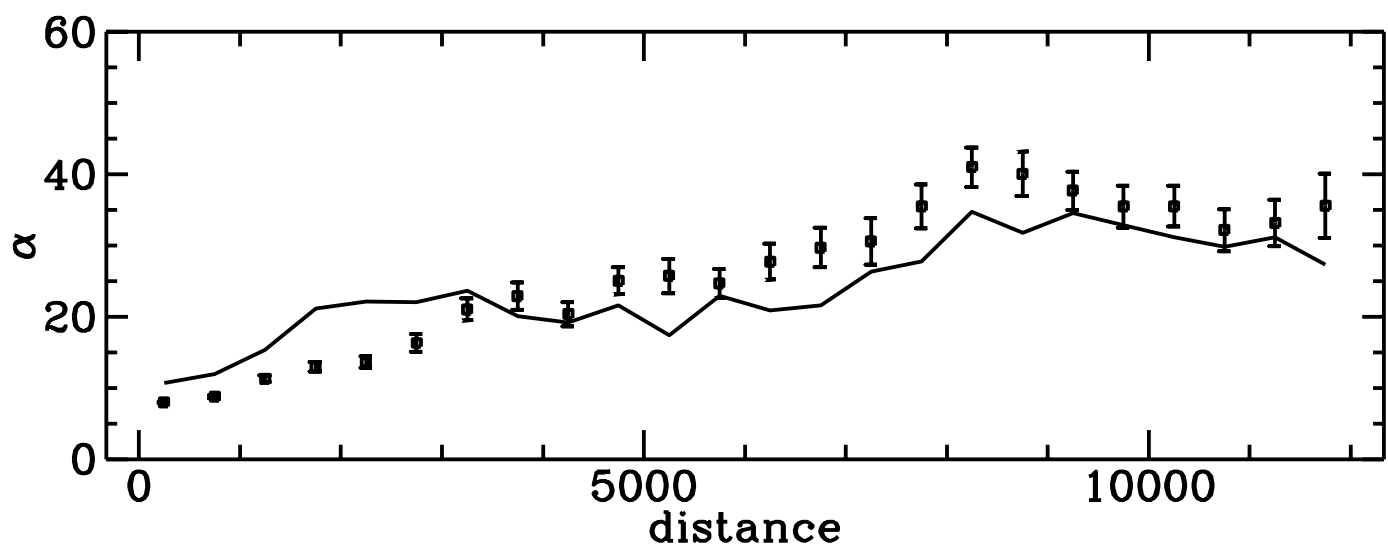
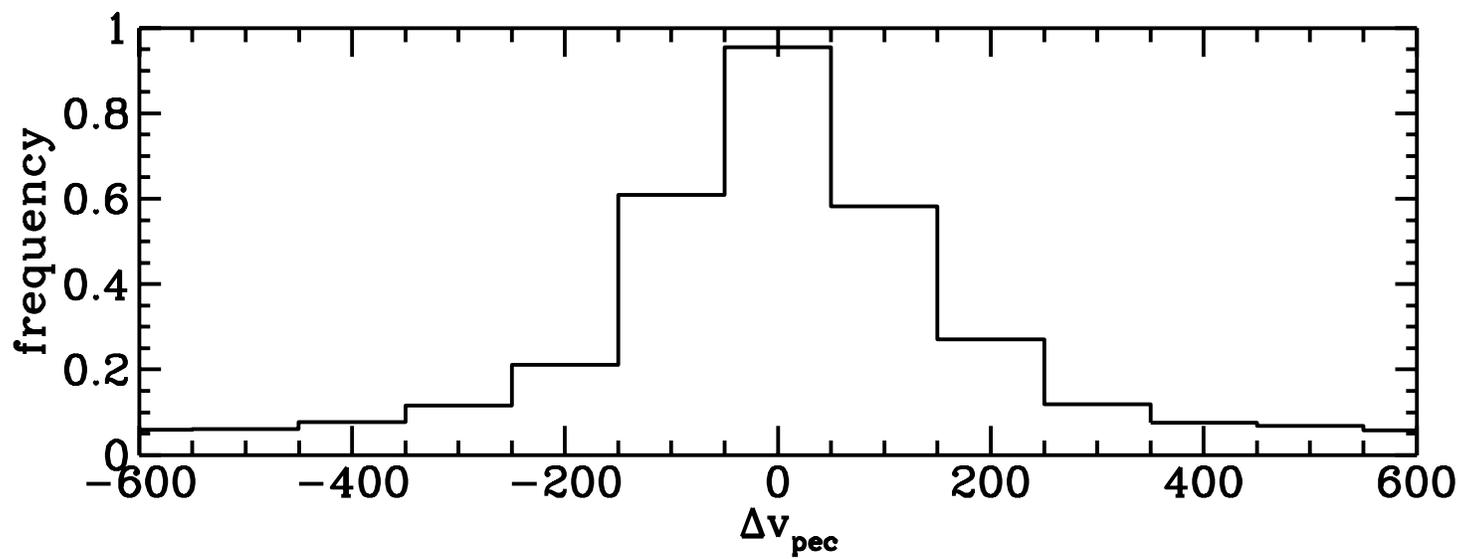

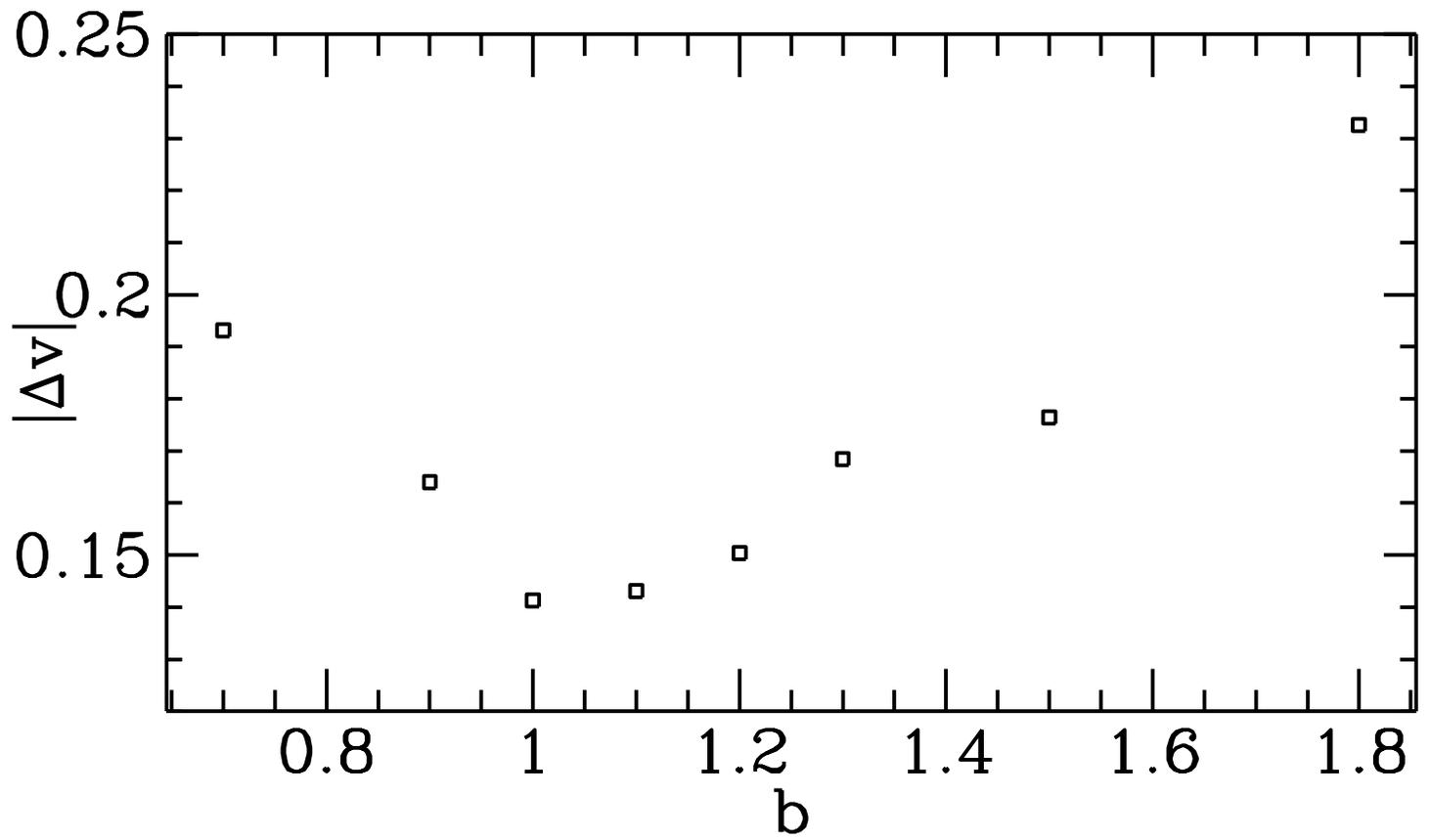

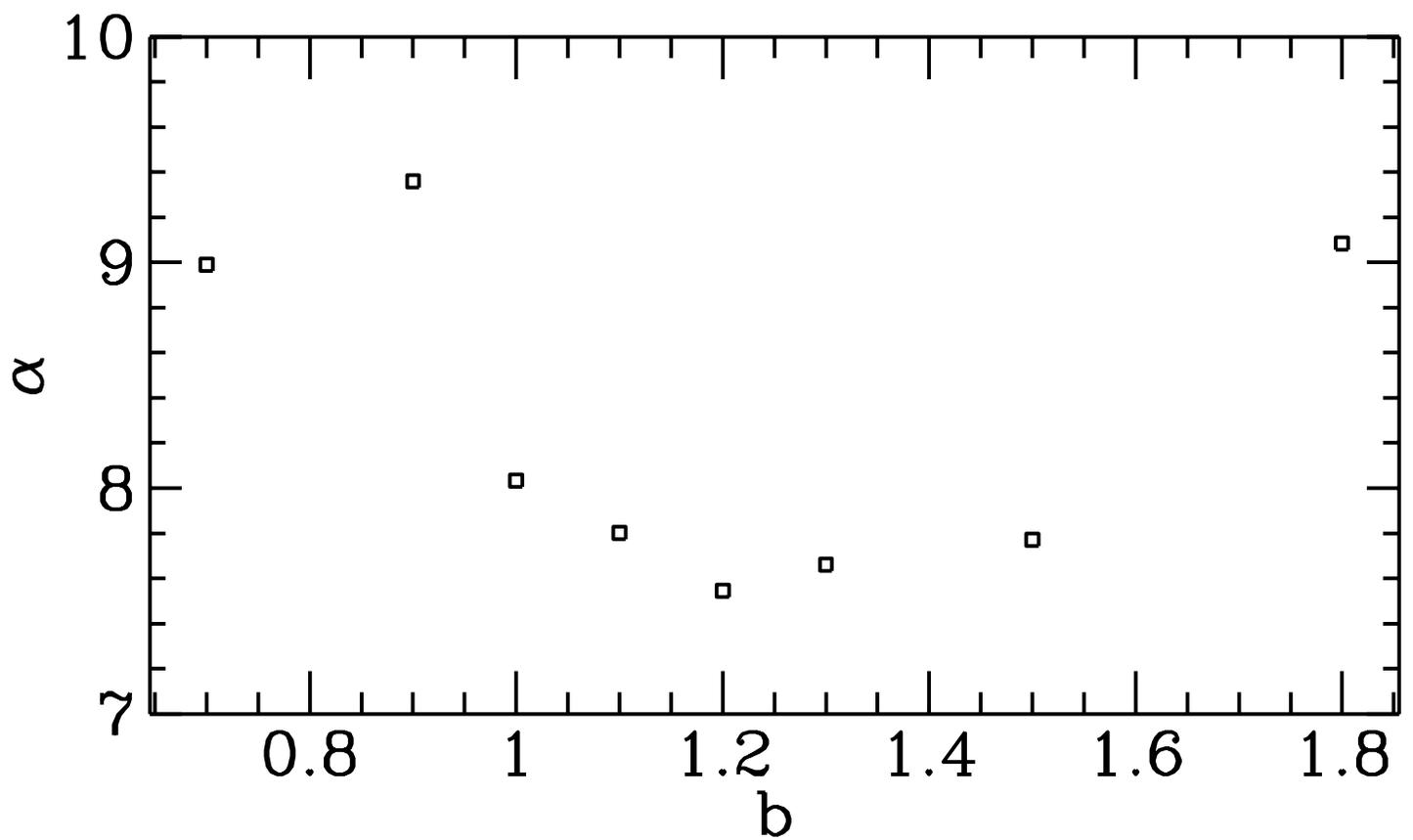

# Testing the Peculiar Velocity Field predicted from Redshift Surveys

Wolfram Freudling[1], Luiz Nicolaci da Costa[2,3,4], Paulo S. Pellegrini [3]

[1] *Space Telescope – European Coordinating Facility, Karl-Schwarzschild-Str. 2, D84748 Garching, Germany*

[2] *European Southern Observatory, Karl-Schwarzschild-Str. 2, D84748 Garching, Germany*

[3] *Observatório Nacional, Rua Gen. José Cristino 77, Rio De Janeiro, Brazil*

[4] *John Simon Guggenheim Fellow*

**ABSTRACT**

The reconstruction of the peculiar velocity field from the 1.936 Jy *IRAS* selected sample of galaxies is compared to a similar reconstruction from an optically selected sample. A general method for combining different samples to reconstruct a self-consistent density and peculiar velocity field is presented. The method is applied to determine how sensitive the derived peculiar velocity field is to the characteristics of the sample used. The possibility that the *IRAS* galaxies do not trace the general galaxy population is explored adopting a simple model of linear biasing between the *IRAS* and optical samples. We find that the velocity fields derived from the two samples are consistent, within the estimated shot noise error, for the case of no relative bias. This result suggests that the predicted peculiar velocity field based on *IRAS* samples is not sensitive to the sampling properties of *IRAS* galaxies. Combined with previous suggestion of a relative biasing of *IRAS* galaxies on small scales (about $5h^{-1}Mpc$), this result suggests scale dependent biasing.

## 1 INTRODUCTION

Under the assumption that galaxies trace the matter distribution, it is possible to reconstruct a self-consistent density and peculiar velocity field from observed redshifts. Comparison of the predicted peculiar velocity field and the available data from various redshift-distance surveys is a powerful approach to the study of large-scale structure, which can be used to test the gravitational instability picture for the growth of structures, to measure the relative distribution of luminous and dark matter and to determine the quantity $\Omega^{0.6}/b$, where $\Omega$ is the density parameter and $b$ is the bias of the galaxy distribution relative to the underlying matter density (e.g. Peebles 1980, Dekel *et al.* 1993 and references therein).

A basic requirement for the self-consistent reconstruction procedure is the availability of a uniform "whole–sky" sample with complete redshift information. This requirement has led to the extensive use of samples drawn from the *IRAS* catalog of galaxies, taking advantage of its wide-angle coverage and uniformity. In particular, this method has been applied by several authors (Yahil 1988, Strauss & Davis 1991, Strauss *et al.* 1992) to a sample of galaxies with complete redshift information to a limiting flux of 1.936 Jy at 60$\mu$m in the *IRAS* PSC database. This sample covers over 87% of the sky.

Unfortunately, the *IRAS* sample is rather dilute, delineating poorly some of the major structures of the galaxy distribution as compared to those observed in optical surveys. In addition, it is well known that the *IRAS* galaxies represent a special class of galaxies and are relatively deficient in high-density regions, as most *IRAS* galaxies are late-type spirals. Some authors have suggested that the discrepancy between the optical and *IRAS* dipoles computed from the two-dimensional distribution could imply that the distribution of *IRAS* galaxies is biased relative to the optical galaxies (e.g. Lahav, Rowan-Robinson & Lynden-Bell 1988).

Here we examine how sensitive the predicted flow is to these effects. This is done by comparing the velocity field obtained using the *IRAS* sample with that derived from a combined *Optical-IRAS* sample. The combined sample used here takes advantage of the denser sampling offered by optical redshift surveys over roughly 50% of the sky, and the wide–angle coverage of the *IRAS* survey. We use this combined sample to investigate the degree of relative bias between the distribution of optical and *IRAS* galaxies. We also use the larger optical sample to estimate the error in the predicted velocity field due to shot noise. The possibility of a relative bias between optical and *IRAS* galaxies has recently been investigated by Strauss *et al.* (1992) by direct comparison of the smoothed density fields obtained from the *IRAS* sample and the CfA1, SSRS and SPS optical samples (Huchra *et al.* 1983, da Costa *et al.* 1991, Dressler 1991). They show that on large scales the two density fields are consistent with them being drawn from the same parent distribution. However, on smaller scales ($\approx$ 500 km s$^{-1}$) there are statistically significant differences between optical and *IRAS* galaxies. Since a range of scales contribute to the peculiar acceleration of a galaxy it is unclear how this may affect the predicted peculiar velocity which is frequently used for comparison with those measured using redshift independent distance estimates. Here we investigate what effect these differences may have on these predictions.

Over the years several attempts have been made to calculate the Local Group velocity based on optical samples using different approaches to treat the low galactic latitude zone (e.g., Davis & Huchra 1982, Pellegrini & da Costa 1990, Hudson 1993). There are several important differences between the current work and previous approaches, which include the optical sample used and a more realistic interpolation across the obscuration zone by utilizing the information available from the *IRAS* catalog. More importantly, this analysis is not restricted to the Local Group but considers the global characteristics of the velocity field.

In section 2 we discuss the details of the method we have utilized for computing the peculiar velocity field, the construction of the combined *Optical-IRAS* sample and how errors were estimated. In section 3 the results of different models are presented together with a discussion of their implications. A brief summary of our main conclusions is presented in section 4.

## 2 THE METHOD

### 2.1 Physical Basis

The measured radial velocity of a galaxy is the sum of its real distance and peculiar velocity

$$v = r + u \cdot \hat{r}, \quad (1)$$

where $v$ is the measured radial velocity in the Local Group rest frame, $r$ is the distance from the sun, $u$ is the peculiar velocity vector, and $\hat{r}$ is the unit vector pointing from the observer along the line of sight. Since only relative distances are relevant to this discussion, we have expressed distances in units of km s$^{-1}$ in order to avoid specifying a value for the Hubble constant.

If the peculiar velocities are gravitationally induced and galaxies trace matter, they will be related to fluctuations of the galaxy distribution. Linear theory of the growth of density perturbation predicts that the peculiar velocity of any given galaxy is

$$u \approx \frac{2g}{3\Omega^{0.4}}, \quad (2)$$

where $g$ is the peculiar acceleration of the galaxy. It is calculated from the potential $\phi$, obtained by numerically solving Poisson's equation

$$\nabla^2 \phi = \frac{3}{2}\Omega(1 + \delta_m) \quad (3)$$

where $\delta_m = \rho_m/<\rho_m> -1$ is the matter density fluctuation; $<>$ indicates an average over the sampled volume.

Once the distribution of galaxies is specified in redshift space, the peculiar acceleration $g$ and the peculiar velocity $u$ can be computed from equations (3) and (2). The peculiar velocity can then be used in equation (1) to improve the estimates of the positions for the galaxies with measured redshift $v$, iteratively. The method breaks down in regions of high density, where there is no unique distance for a given redshift. In these regions, we follow method 2 of Yahil et al (1991) and adopt the distance which is the average of the minimum distance where the computed redshift is 200 km s$^{-1}$ less than the observed redshift and the maximum distance where the computed redshift is 200 km s$^{-1}$ more than the observed redshift. For every galaxy, the peculiar velocity was computed at 100 distances along the line of sight and for each such point the corresponding redshift. From the resulting redshift–distance relation, galaxies which need this special treatment were identified and their distance computed. A different strategy to treat such regions has been discussed by Freudling *et al.* (1991).

Since the computation of the peculiar velocities relies on a linear approximation, it cannot be applied within clusters. Therefore, clusters from the list given by Yahil *et al.* (1991) have been collapsed in redshift space by setting the initial distances and positions of the cluster galaxies to that of the center of the cluster.

While the physical basis of the algorithm used here to compute the accelerations **g** is identical to the one used by Yahil *et al.* (1991), the numerical approach is different (Freudling *et al.* 1991). Instead of adding the contribution of individual galaxies, the gravitational acceleration **g** is computed from the smoothed density field with a particle-mesh code via equation (3). The density field is calculated on a grid with a 500 km s$^{-1}$ cell size. The total dimension of the grid is $64^3$, i.e. the sphere with radius 8000km s$^{-1}$ on which $\delta$ is not equal to zero occupies only a fraction of the cube. These dimensions were chosen so that galaxies do not interact with the mirror image created by the periodic boundary conditions of the particle-mesh code (see Freudling et al., 1991). The weight of each galaxy is assigned to the grid so that the resulting smooth distribution does not introduce any dipole or quadrupole moments (TSC assignment scheme). The weight assigned to galaxies is the inverse of the appropriate selection function. Davis & Huchra (1982) have shown that this assignment is virtually identical to the minimum variance estimator. Finally, the relative density fluctuation $\delta$ is computed for grid points in the surveyed region ($v < 8000$ km s$^{-1}$). The advantage of this algorithm is that it provides direct access to the $\delta$ used as input to the linear perturbation theory. We take advantage of this for the interpolation over the galactic plane and for the implementation of the bias of the *IRAS* sample relative to the optical sample. In principle, a simple nonlinear correction, as suggested by Nusser et al. (1992), could also be easily built into the model.

## 2.2 Construction of the Density Field

### 2.2.1 Samples

In this paper we have considered two samples: the first is the *IRAS* 1.936 Jy sample of Strauss *et al.* (1992); the second one is a combination of the optical sample available in the galactic caps and the part of the *IRAS* sample located at low galactic latitudes, which are not covered by the former.

The optical sample consists of galaxies brighter than $m_{B(0)}=14.5$. In the northern galactic cap (b$\geq 40°$) we use the CfA1 sample (Huchra *et al.* 1983). In the southern galactic cap (b $\leq$ -30°) we use the so-called SGC sample assembled by Pellegrini *et al.* (1990) from different catalogs (CGCG, ESO, MCG) with redshifts taken primarily from the SSRS (da Costa *et al.* 1991) and the equatorial survey (Huchra *et al.* 1993). This combined catalog has been shown to be fairly homogeneous and its magnitude scale to be statistically consistent with the B(0)-Zwicky system adopted in the CfA1 survey. Recently, improvements have been made to this catalog south of $\delta = -17.5°$ and below $b = -30°$ by adopting the $B_T$ magnitudes listed in the ESO photometric catalog (Lauberts & Valentijn 1989), converted to the B(0) system, instead of using magnitudes estimated from diameter-magnitude relations. A more detailed discussion on the accuracy of the resulting magnitudes has been presented by Alonso *et al.* (1993).

### 2.2.2 Density Field for the Optical-IRAS Sample

If only one sample is considered the density fluctuation $\delta$ can be computed directly from the number counts per volume $\rho$ as $\delta = \rho/<\rho> -1$. This implicitly assumes that the surveyed volume represents a fair sample of the universe and that the average density fluctuation $<\delta>$ is zero within this volume. In order to impose the same assumption that the total volume averaged $\delta$ of the combined *Optical-IRAS* sample equals zero, the differences in the mean density between the two distinct samples must be taken into account.

This can be achieved in the following way. To compute $\delta$ for our combined sample from the number densities $\rho_o$ and $\rho_i$ in the optical and *IRAS* part of the sample, respectively, we use the substantial overlap between these samples in the galactic caps ($b < -30°$ or $b > 40°$) to define

$$\rho = \begin{cases} \rho_o & \text{in caps} \\ \rho_i \frac{<\rho_o>_c}{<\rho_i>_c} & \text{in plane} \end{cases} \quad (4)$$

where $<>_c$ is the average of a quantity in the galactic caps. With this expression, the density $\rho$ is computed over the entire volume, and its average used to determine the density fluctuation

$$\delta = \begin{cases} \rho_o/<\rho> -1 & \text{in caps} \\ \frac{\rho_i}{<\rho>}\frac{<\rho_o>_c}{<\rho_i>_c} - 1 & \text{in plane} \end{cases} \quad (5)$$

In order to account for the possibility of a relative bias between the *IRAS* and optical samples we introduce the relative biasing factor $b_r = \delta_{opt}/\delta_{iras}$, where $\delta_{iras}$ and $\delta_{opt}$ are the density fluctuations of the *IRAS* and optical samples, respectively. Here we have adopted a simple linear biasing scheme between these samples, assuming that $b_r$ is a constant independent of position. For models which utilize the combined sample, equation (5) changes to

$$\delta = \begin{cases} \rho_o/<\rho'> -1 & \text{in caps} \\ b_r \left( \frac{\rho_i}{<\rho'>}\frac{<\rho_o>_c}{<\rho_i>_c} - 1 \right) & \text{in plane} \end{cases} \quad (6)$$

where the density $<\rho'>$ is adjusted to make $<\delta>=0$ as before. In the next section we use this equation to compute $\delta$ for different values of $b_r$.

For grid points at high redshifts, beyond the velocity limit imposed on our samples, the density fluctuation $\delta$ was set to zero. At low galactic latitudes, we follow Yahil *et al.* (1991) and interpolate over the region $|b| < 5°$ by setting the density equal to the average of the density in the 2 strips $5° < |b| < 15°$ in 1000 km s$^{-1}$ redshift bins and 10° longitude bins. This interpolation can again contribute to the mass contained within the sphere defined by the maximum redshift, because the average density fluctuation $\delta$ in the interpolation zone is not necessarily zero. This contribution has to be taken into account when the average density $<\rho'>$ is computed. Once the initial density field has been constructed the system is allowed to evolve as outlined below.

### 2.3 Implementation

The complete computational procedure was as follows. As a preparation for the iterative part, the observed redshifts of the galaxies were corrected to that relative to the center of the Local Group, which was assumed to move toward $l = 270°$, $b = 0°$ with 300 km s$^{-1}$. The initial distance of all field galaxies was set to their corrected redshifts, and clusters collapsed as described above.

Each iteration started with the computation of the selection function, following the procedure adopted by Davis & Huchra (1982). For the combined *Optical-IRAS* samples, the function for the *IRAS* part of the sample was assumed to be that derived from the corresponding all-sky *IRAS* model in order to reduce the uncertainty due to the smaller sky coverage.

To improve the convergence of the procedure, the density parameter $\Omega$ was set to a fraction of its final value in the first iteration, and gradually increased to its final value, always taken to be unity. This has the effect that the density field, and therefore the gravitational field, changes only slightly between iterations.

The next step is to construct the density field from the discrete distribution of the galaxies. This includes the distribution of the galaxy weights to the grid, biasing and interpolation over the galactic planes which has been discussed in detail in previous section. Subsequently, equation (3) is solved to compute the gravitational field and peculiar velocities by equating the matter density fluctuation $\delta_m$ to the optical galaxy density fluctuation given by equation (6). Together with our assumption that $\Omega = 1$ this is equivalent to models with $\beta = \Omega^{0.6}/b = 1$, where $b$ is the bias factor of the optical sample relative to the matter distribution.

The distances of the galaxies is adjusted for the next iteration to be the average between their previous distance and the distance which produces the observed redshift given the computed peculiar velocity, or using the procedure described above where necessary to avoid ambiguities.

The rms of this distance correction as well as the maximum adjustment was recorded, and the next iteration started with the new distances. In typical runs, the models converged to an rms of about 10 km s$^{-1}$ in 5 iterations after the final set of parameters was reached.

A preliminary value for the relative weight between the *IRAS* and the optical part of the sample $<\rho_o>/<\rho_i>$ was computed from the real space distributions for the *IRAS* and optical models restricted to the galactic caps (see discussion below) and was found to be 2.40. Subsequently, this value was used as input for a preliminary version of the whole-sky, unbiased *IRAS* and *Optical-IRAS* models, and was recomputed from the real space distribution of these models, yielding $<\rho_o>/<\rho_i> = 2.45$. This value was adopted in all other models. This factor computed separately for the northern and southern cap resulted in values which differ from each other by 22%.

### 2.4  Error Estimates

In order to estimate the shot noise typical of *IRAS*-like samples we have used the larger *Optical-IRAS* sample in the following manner. Since most of the contribution to the shot noise comes from large distances, we created subsets of the large optical sample which had approximately the same number density as the *IRAS* sample near our velocity limit of 8000 km s$^{-1}$. The optical sample at 8000 km s$^{-1}$ contains about twice as many galaxies per volume as the *IRAS* sample. Therefore we randomly divided the optical sample into two. The purpose of creating these diluted samples was to examine the difference between the *IRAS* and the *Optical-IRAS* velocity models simply due to the sampling, i.e. the shot noise. These diluted samples contain about twice the number of galaxies per solid angle as the *IRAS* sample in the galactic caps, which reflects the different radial distribution of the two samples. These optical subsamples were then combined with the *IRAS* sample restricted to the galactic plane in the same way as it was done for the complete optical sample.

The difference in the predicted peculiar velocities of the two diluted samples was used as an empirical estimate of the shot noise as a function of distance. Local Group velocities derived from the models without biasing are given in Table 1, where the number of *IRAS* and optical galaxies in the diluted samples and the direction and amplitude of the Local Group velocity vector predicted from these models are given. Figure 1 quantifies the difference between those models as a function of distance. The upper panel shows the average difference of the absolute value of the peculiar velocities as a function of distance. This comparison includes both galaxies within and outside of the tripled valued regions. Galaxies have been binned in 500 km s$^{-1}$ distance bins. The error bars show the $rms/\sqrt{N}$ in each bin, where $N$ is the number of galaxies in that bin. The middle panel shows the angular difference between the velocity vectors in both models. The binning and error bars are computed as in the upper panel. The lower panel shows the distribution of the differences in the x, y and z components of the peculiar velocities. Note that this distribution is not necessarily expected to be symmetrical because differences are computed at the positions of galaxies, which might sample the velocity field preferentially where differences are negative or positive.

Errors in the amplitude are less than 200 km s$^{-1}$, and angular differences are in the range of 20° to 30°. As can be seen in figure 1, the shot noise is small at small distances. This is because velocities are relative to the Local Group rest frame and therefore by definition zero at the origin in all models, and the sampling is densest at small distances. The shot noise increases with distance as the sampling becomes more dilute. At distances greater than the velocity limit of the sample (8000 km s$^{-1}$), all contributions to the gravitational acceleration come from the inner volume and therefore the shot noise decreases as this limit is approached. Since all peculiar velocities are computed in the Local Group rest frame, predicted peculiar velocities in each model converge towards the projection of the predicted Local Group velocity for galaxies at large distances. Therefore, for very large distances, the differences between any two models approach asymptotically the difference in the estimated Local Group velocity for each sample.

This comparison was used as an estimate of expected differences due to shot noise between the full *Optical-IRAS* model and the *IRAS* model. However, this error estimate was based on two *IRAS*-like diluted samples. Due to the denser sampling of the optical distribution, this will be an overestimate. To take this into account, we divide the original error estimate by $\sqrt{2}$, which will be used as a lower bound to our error estimate, whereas the original error will be used as an upper bound.

## 3 RESULTS

First, to test our procedure, the velocity field was computed using only the *IRAS* sample, limited to a redshift of 8000 km s$^{-1}$ and taking $b_r = 1$. For this case the predicted velocity field, in cartesian supergalactic coordinates, is shown in figure 2. In this representation the SGX-axis points to SGL=0° and the SGY-axis points nearly in the direction of the north galactic pole. Each row contains three parallel and non-overlapping 3000 km s$^{-1}$ thick slabs, with the middle slab centered at the origin. In these figures the SGY=0 plane is almost coincident with the galactic plane while the SGZ=0 plane is coincident with the supergalactic plane. The unit throughout is km s$^{-1}$ and the amplitude of the vectors are in the same units as the grid. The data is presented in the Local Group rest frame. In each panel the vectors represent the projection onto the corresponding plane of the predicted three-dimensional velocity vector. The velocity field of figure 2 shows good agreement with those of Yahil (1988), who plotted galaxies within a 45° wedge about the Supergalactic plane, confirming that the results are insensitive to the details of the method used. This can be seen, for instance, in the special case of the induced velocity of the Local Group for $b_r = 1$, given in Table 2. The columns are as follows: column (1) the sample used; column (2) the bias factor; columns (3) and (4) the number of *IRAS* and optical galaxies in the sample; column (5) the number of galaxies in triple valued regions in the model, columns (6) and (7) the galactic longitude and latitude of the direction of the Local Group velocity vector; column (8) the amplitude of the Local Group velocity and column (9) the angle between the computed Local Group velocity vector to the cosmic microwave background dipole (Smoot *et al.* 1992). The predicted Local Group velocity for the unbiased model is within 4 degrees in direction and 5% in amplitude of that obtained by Strauss *et al.* (1992) for a similar model.

Next, several models using the combined *Optical-IRAS* samples were computed, for values of relative biasing factor $b_r$ in the range 0.7-1.8. For comparison, similar models with the *IRAS* sample were computed. These *IRAS* models, with different values of $b_r$, are intended to simulate whole-sky optical samples within the framework of a linear biasing model. The velocity field derived for $b_r = 1$ is illustrated in figure 3. The velocity vectors are only shown at the position of the *IRAS* galaxies. The similarities between the *IRAS* and *Optical-IRAS* samples are striking. In order to assess their differences, the velocity fields derived from the *Optical-IRAS* sample were subtracted from those derived by the corresponding biased *IRAS* models. The residual fields are shown and figures 4 and 5. Comparison of these figures clearly illustrates that the difference between $b_r = 1$ *IRAS* and *Optical-IRAS* models are much smaller than between the $b_r = 1.3$ models. In figures 6 and 7 the *rms* differences as a function of distance were compared with the estimated errors obtained from the diluted samples. As can be seen from the examination of these figures for the case $b_r \approx 1$ the departures are small while for $b_r = 1.3$ significant departures can be observed out to 4000 km s$^{-1}$.

Instead of looking at the global velocity field we can focus our attention on the peculiar velocity induced to the Local Group. In Table 2 we list the values obtained for the Local Group velocity for some of the different models described above, while in figure 8 we show the vector difference of the Local Group velocity predicted by the *Optical-IRAS* and *IRAS* sample as a function of the relative biasing factor. It is interesting to note that the Local Group velocity vector for the *Optical-IRAS* and *IRAS* models with $b_r \approx 1$ is essentially the same. This result could indicate that no linear relative bias between optical and *IRAS* is required. Another possibility is that the velocity field even at large distances from the galactic plane is primarily determined by structures near it, where our combined *Optical-IRAS* sample is by construction identical to the *IRAS* sample. This explanation would be surprising since major structures such as the supergalactic plane, Virgo and the Southern Wall are located at high galactic latitudes. We tested this hypothesis by running the same models with the density fluctuations $\delta$ in the galactic plane set to zero. The *Optical-IRAS* models are thus reduced to the optical sample only and the *IRAS* sample is restricted to the same volume. The resulting Local Group velocities for $b_r = 1$ are also given in Table 2. The agreement between the Local Group velocity vector derived from these two models is again excellent both in direction and amplitude, although they are now constructed from completely different galaxy samples. We conclude that any linear relative biasing between optical and *IRAS* galaxies is small.

## 4  CONCLUSIONS

In this paper we have compared the peculiar velocity field predicted from an *IRAS* and an optical sample in order to directly test the effects that diluteness and undersampling of high-density regions may have on the computed velocity field from the *IRAS* catalog of galaxies. Since the available optical samples have limited sky coverage, we have adopted a combined *Optical-IRAS* catalog and modeled the differences between the density fluctuations in terms of a linear biasing model.

We find that the predicted flow is similar, within the estimated shot noise, if the relative bias is small. This result suggests that the relative biasing between the *IRAS* sample and the optical sample is small on scales which induce the peculiar velocities. We find that this is true even when the samples are restricted to the galactic caps, in which case we are directly comparing independent optical and *IRAS* samples. These results suggest that the predicted velocity field derived from the *IRAS* catalog is robust and that *IRAS* galaxies are fair tracers of the galaxy population. The good agreement of the velocity fields from these two samples also confirm, a posteriori, our original claim that it is possible to construct a fairly homogeneous "all-sky" optical sample.

## Acknowledgements

We would like to thank C. Willmer for his collaboration in the preparation of the optical sample and M. Strauss for making available the *IRAS* sample. LNdC also thanks the hospitality of the Harvard-Smithsonian Center for Astrophysics where a substantial part of this work was carried out.

**Table 1.** Diluted Models.

| $n_{iras}$ | $n_{opt}$ | $l_{lg}$ (°) | $b_{lg}$ (°) | $v_{lg}$ (km s$^{-1}$) |
|---|---|---|---|---|
| 970 | 1945 | 241 | 38 | 734 |
| 967 | 1937 | 246 | 32 | 749 |

**Table 2.** Predicted Local Group Velocities.

| sample | $b_r$ | $n_{iras}$ | $n_{opt}$ | $n_{trip}$ | $l_{lg}$ (°) | $b_{lg}$ (°) | $v_{lg}$ (km s$^{-1}$) | $\theta$ (°) |
|---|---|---|---|---|---|---|---|---|
| *IRAS* | 0.7 | 1897 | 0 | 7 | 251 | 36 | 574 | 23 |
| *IRAS* | 1.0 | 1893 | 0 | 235 | 252 | 34 | 726 | 21 |
| *IRAS* | 1.3 | 1884 | 0 | 620 | 253 | 35 | 881 | 21 |
| *Optical-IRAS* | 0.7 | 970 | 3873 | 99 | 241 | 39 | 646 | 31 |
| *Optical-IRAS* | 1.0 | 969 | 3873 | 244 | 242 | 34 | 719 | 29 |
| *Optical-IRAS* | 1.3 | 961 | 3932 | 499 | 249 | 35 | 776 | 24 |
| *IRAS* in caps | 1.0 | 928 | 0 | | 232 | 65 | 509 | 44 |
| *optical* in caps | 1.0 | 0 | 3851 | | 207 | 59 | 501 | 54 |

**Figure Captions**

Figure 1- Differences in the velocities fields of the two diluted models. The upper panel shows the differences in the absolute value of the predicted velocities as a function of distance. Galaxies have been binned into 500 km s$^{-1}$ bins according to the average distance in the two models. The error bars shown are the rms divided by the square root of the number of galaxies in each bin. The middle panel shows the differences in the direction of the velocity vectors in degrees, again binned into 500 km s$^{-1}$ bins. The error bars are computed as in the previous panel. In the lower panel the distribution of the differences in x, y and z direction are shown.

Figure 2- Predicted velocity field computed from the density field of *IRAS* galaxies. Each panel shows the velocity components, projected onto the indicated planes, of galaxies within 3000 km s$^{-1}$ slabs. To avoid overcrowding, velocity vectors are shown only for a sample semi-volume limited at 4000 km s$^{-1}$. Distances and velocities are both in km s$^{-1}$. Panels are ordered from the top-left (a) to the bottom-right (i). We use supergalactic coordinates as explained in the text.

Figure 3- Same as figure 2, showing the predicted velocity field, at the position of the *IRAS* galaxies computed from the density field of the *Optical-IRAS* sample and $b_r = 1$.

Figure 4- Same as figure 2, but showing the difference of the predicted velocity, at the position of *IRAS* galaxies, computed from the density field of the *Optical-IRAS* and *IRAS* sample. The relative biasing $b_r$ was 1 in both models.

Figure 5- Same as figure 4, for $b_r = 1.3$.

Figure 6- Differences in the velocities fields between the *Optical-IRAS* and *IRAS* models for the case $b_r = 1$. The meaning of the three panels are described in the caption to figure 1. The solid and dashed lines represent the upper and lower bounds of our error estimate.

Figure 7- Same as figure 6 for the case $b_r = 1.3$.

Figure 8- Differences in the Local Group velocity vectors as a function of the relative biasing factor. The upper panel shows the fractional change of the amplitude, while the lower panel shows the difference in the angular direction.